\documentclass[fleqn,usenatbib,useAMS]{mnras}

\usepackage[utf8]{inputenc}		%
\usepackage[T1]{fontenc}		%
\usepackage{ae,aecompl}         %
\usepackage{graphicx}			%
\usepackage{amssymb} 			% Symbols
\usepackage{amsmath} 			% Symbols
\usepackage{wasysym}
\usepackage{url}
\usepackage{multicol}           % Multi-column entries in tables
\usepackage{bm}		            % Bold maths symbols, including upright Greek
\usepackage{pdflscape}	        % Landscape pages

\title{Towards a Classification Scheme for the Rocky Planets based on Equilibrium Thermodynamic Considerations}

\author[O. Bertolami \& F. Francisco]{Orfeu Bertolami$^{1}$ \& Frederico Francisco$^{1}$%
\\
$^{1}$Departamento de Física e Astronomia and Centro de Física do Porto, Faculdade de Ciências, Universidade do Porto,\\ Rua do Campo Alegre 687, 4169-007 Porto, Portugal}

\date{Last updated \today; in original form \today}

\pubyear{2021}

\begin{document}

\maketitle

\begin{abstract}
\noindent
A classification scheme for rocky planets is proposed, based on a description of the Earth System in terms of the Landau-Ginzburg Theory of phase transitions. Three major equilibrium states can be identified and the associated planetary states or phases are: Earth-like Holocene state;  hot Venus-like state; cold Mars-like state. The scheme is based on an approach proposed to understand the Earth transition from the Holocene to the Anthropocene, driven by the impact of the human action on the Earth System. In the present work we identity the natural conditions that cause transformations on the planets forcing them into one of the states identified above. We discuss how the parameters that describe these transformations can be related with exoplanets observables. In analysing the relevant physical parameters, we were stroke by the similarities between Earth and Venus, and how likely is that the Anthropocene transition may lead to hot-house Earth scenario.  
\end{abstract}


%-----------------------------------------------------------------------------%
%-- Body ---------------------------------------------------------------------%
%-----------------------------------------------------------------------------%

\section{Introduction}
\label{sec:introduction}

In this work, we propose a classification scheme for rocky planets inspired on a description of the Earth System (ES) in terms of the Landau-Ginzburg Theory of phase transitions \cite{OB-FF1,OB-FF2,OB-FF3,OB-FF4}.

Over its evolution, the ES, the system resulting from the integration and interaction between the atmosphere, biosphere, cryosphere, hydrosphere and upper lithosphere, has been driven predominantly by astronomical and geophysical forces. Generalising this description to other rocky planets seems quite natural, even though some components for a detailed description are still missing. Work in this direction is quite timely given that, as we write in late March 2022, 5005 exoplanets have been discovered 
(https://exoplanetarchive.ipac.caltech.edu) 
and thousands of millions are expected to exist in our galaxy (see, for instance, Refs. \cite{Lee,Howell} for reviews). 

We thus argue that rocky planets in or nearby a habitable zone should be described in terms of a subset of the components and features encountered on Earth. Their evolution and equilibrium states can be understood as a result of transitions similar to phase transitions. 

Of course, Earth is a rather special case given the importance of its biosphere and its interactions with all other subsystems, and also by the fact that over the last few decades human activities have been intensified to the point of given origin to a new geological epoch, the Anthropocene \cite{Crutzen:2000,Crutzen:2002,Steffen:2007,Steffen:2011,Waters:2016}. Furthermore, Earth is the touching stone of any classification scheme of rocky planets given the depth of our knowledge about its composition, structure and evolution. Our classification scheme is no exception in this respect as it springs from previous studies aimed to understand Earth's transition from the Holocene to the Anthropocene.  

The description we propose assumes that the equilibrium state of a planet depends on physical parameters such as its surface temperature, the energy received by its star, its efficiency to clean its orbit, its volcanic activity and the presence of plate tectonics activity. The property of cleaning its orbit from debris is usually captured either by the Stern-Levinson parameter \cite{SL2002}, $\Lambda$, or by the Soter's parameter \cite{Soter2006}, $\mu$, or the Margot's parameter \cite{Margot2015}, $\Pi$, which will be defined below. Volcanic activity in the Solar System, found in a wide range of planets and large satellites, in the latter often in the form of cryovolcanoes, is clearly a quite relevant parameter as it plays a crucial role in releasing intrinsic heat and to expel $\rm CO_2$ into the atmosphere. These factors are relevant to heat up and stabilise the temperature of a planet and to drive transformations of its thermodynamical state. In what concerns plate tectonics, although in the Solar System it is only found on Earth and possibly on Venus in the past, it is clearly a quite relevant factor: tectonic plates can be the cause of subduction, which induces volcanic activity and the release of $CO_2$. It has been argued that the relevant parameter to gauge if a planet, in its equilibrium state, is monoplate or whether this initial single plate cracks, is its mass. Super-Earth planets are likely to host plate tectonics as when formed they retained sufficient heat to induce transformations on its original plate \cite{Valencia2007}. However, it has been argued, in this respect, that in order to establish the Earth-like nature of a planet requires a composition analysis in which the critical parameter is the criterion, $Si/Fe>0.9$ \cite{Unterborn2016}, where for Earth, $Si/Fe=1.0$ (and for the Sun $Si/Fe=1.19)$. The presence of water and oceans can be a favourable condition for the existence of plate tectonics, which are also believed to be associated for the emergence and maintenance of life.

We believe that these features are sufficiently general to encompass the great majority of rocky exoplanets and, in any case, as we shall see, it is rather easy to incorporate more parameters in our description. In what follows we shall assume that planets, since their origin, have undergone transformations that led them to eventually settle into equilibrium states. In its simplest version, our model predicts three equilibrium states: Earth-like Holocene state; hot Venus-like state; cold Mars-like state. Of course, significant variation of the relevant parameters can cause transformations that force a planet to move from one equilibrium state into another. The generality of the assumptions we have adopted to specify the three major classes of planets imply that we expect that rocky planets can either spend most of their existence in one of the three possible states or they can evolve from one category into another depending on the nature or the intensity of transformations they undergo through. Clearly, Mars is a prototypical example of this type of dramatic change as the abundance of morphological features suggesting the presence of water in its past, indicates that it could even have sheltered life, but it has clearly lost this capacity.  

We point out that our considerations do not concern Mercury-like planets as their dynamics is strongly conditioned by the proximity of their orbit to the mother star and the amount of radiation it receives, which tends to overwhelm their specific intrinsic features. Indeed, a recent study shows that rocky super-Mercury planets stand out by their higher density, but also by their higher iron content, very much in opposition to other rocky-type planets \footnote{Even though their are considered teluric-type planets, that is, planets rich in silicates and metals, distributed in a differentiated structure, with a metal core, and rocky surface} whose composition is well correlated with the composition of their host star \cite{Adibekyan}.

Notice that our approach differs, for instance, from the one proposed in Ref. \cite{FrankAlbertiKleidon2017} based also on thermodynamic arguments, but whose emphasis is non-equilibrium thermodynamics of coupled planetary systems and is predominantly concerned with transformations driven by the biosphere, that is by life and also intelligence. Similarities with the well known Kardashev classification \cite{Kardashev1964} are evident. Our approach can, of course, deal with out-equilibrium transitions between equilibrium states, as it does, for instance, for Earth's Anthropocene, however it seeks a broader look to the planets and their stability states driven by physical parameters assuming that although the biosphere is a determinant component of a planetary system, the vast majority of planets are not endowed at all with this component or when they do this is relatively incipient.    

This work is organised as follows. In the next section, we shall present our model. In section 3 we shall discuss the relevant parameters of our classification scheme and in section 4 we discuss how the parameters of our classification can be related with exoplanets observables. I section 5, we present our conclusions and a general discussion of the proposed classification scheme and its implications, for instance, for terraforming transformations.  

%-----------------------------------------------------------------------------%
%-----------------------------------------------------------------------------%

\section{Planetary Phase-transition Model}
\label{sec:model}

%-----------------------------------------------------------------------------%

\subsection{Landau-Ginzburg Model of the Holocene-Anthropocene Phase Transition}

Our starting assumption is that the equilibrium states of a planet compatible with a given thermodynamical state are associated to different phases and that the transitions between equilibrium states are similar to phase transitions that can be described by the Landau-Ginzburg theory (LGT) \cite{Landau:1937,Ginzburg:1960}, a theory that is known to account for  phase transitions in a wide range of systems in condensed matter, material science, quantum field theory and cosmology. 

The Landau-Ginzburg theory (LGT) describes phase transitions of physical systems in terms of their free energy through an expansion in terms of an order parameter, $\psi$. This parameter is a function of some state variable of the system, such as temperature or magnetisation. The phase transition occurs between different phases corresponding to different minima of the free energy. It is usual in this context to refer to a ``symmetric'' phase corresponding to few possible macroscopic equilibrium states, while a ``asymmetric" phase has more equilibrium states  \cite{Landau:1937,Ginzburg:1960}. 

A rather prototypical situation is that of a continuous phase transition at a temperature $T_{\rm c}$ from a higher temperature state to a lower temperature state, where initially the order parameter $\psi(T)$ is zero for $T > T_{\rm c}$ and is non-vanishing for $T < T_{\rm c}$, where $T_{\rm c}$ is a critical temperature. In the LGT, the free energy of the system is given by an analytic function of the order parameter. The simplest form that yields a phase transition is
\begin{equation}
    F(\eta) = F_0 + a(\eta) \psi^2 + b(\eta) \psi^4 + \ldots,
    \label{eq:FreeEnergy}
\end{equation}
where $\eta$ denotes the set of parameters that can change the state of the system, $F_0 = F(\eta_0)$ is a constant and $a(\eta)$, $b(\eta)$ are analytic functions of $\eta$,
\begin{equation}
{\setlength\arraycolsep{2pt}
\begin{array}{r c l}
a(\eta) &=& a_0 \eta + a_1 \eta^2 + \ldots,\\
b(\eta) &=& b_0 + b_1 \eta + \ldots, \\
\end{array}}
\label{eq:coefficients}
\end{equation}
where $a_0$, $b_0$, are positive constants and all coefficients have units of energy. The coefficients $a_0,\ldots,b_0,\ldots$ should, in principle, be observables and can be fitted from data.

This description is valid for a spatially homogeneous systems. However, the theory remains valid for non-homogenous systems by adding terms depending on the gradient, $|\nabla\psi|^2$, and the Laplacian, $\nabla^2\psi$, of the order parameter. For simplicity, we shall assume no spatial dependence of $\psi$, in effect, not accounting for local variations. 

When using this framework to describe the Earth System, the order parameter $\psi$ is the reduced global average temperature and $\eta$ are the Astronomical, Geophysical and Internal dynamics drivers of the Earth System.

As mentioned, at the minimum of the free energy function, one can find the equilibrium states of the system. Indeed, considering Eq.\,\ref{eq:FreeEnergy}, one can find minima of the order parameter, corresponding to the equilibrium states, denoted by $\langle\psi\rangle$, for the distinct phases:
\begin{equation}
\frac{dF}{d \psi} = 0 \Rightarrow \left\{
\begin{array}{l l}
        \langle\psi\rangle = 0, 	& \eta > 0 \\
        \langle\psi\rangle= \pm \left(-\frac{a(\eta)}{2 b(\eta)} \right)^\frac{1}{2}, 	& \eta < 0 \\
    \end{array}
    \right. ,
\label{eq:Minima}
\end{equation}

\begin{figure}
    \centering
    \includegraphics[width=\columnwidth]{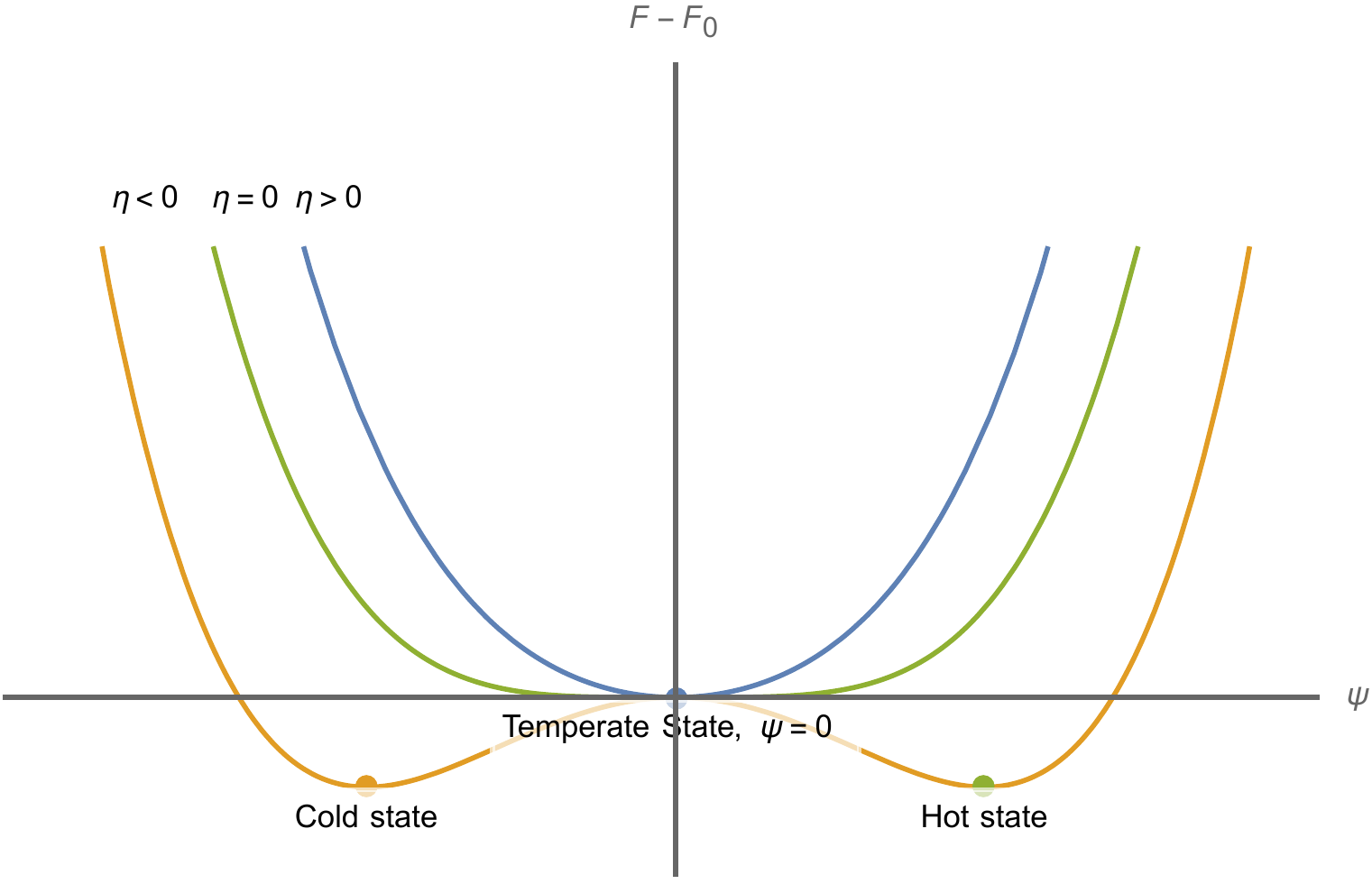}
    \caption{Free energy as a function of the order parameter driven only by astronomical, geophysical and internal dynamics expressed in terms of the variable $\eta=(A,G,I)$. The Earth-like Holocene stability is associated with the phase $\langle\psi\rangle = 0$, for which $\eta > 0$.}
    \label{fig:ESLGT}
\end{figure}

As shown in Fig.\,\ref{fig:ESLGT}, the two phases are clearly distinguishable by the position of their minima. In the so-called ``symmetric'' phase, with $\eta > 0$, there is a single stable minimum for a temperate state at $\langle\psi\rangle = 0$. We hypothesise that this phase models the stability of Earth's Holocene. The other phase is an ``asymmetric'' phase, when $\eta < 0$. In this phase, there are two stable minima, one at a hot state and the other at a cold state. In the case of Earth, there have been numerous periods of hotter climate and of glaciation that could be described by these states. When there is a transition from $\eta > 0$ to $\eta < 0$, the system starts from the critical state $\psi = 0$ and goes either to a hotter or to a cooler equilibrium state.

%-----------------------------------------------------------------------------%

\subsection{Generalisation to Generic Rocky Planets}

In previous studies \cite{OB-FF1,OB-FF2,OB-FF3,OB-FF4}, in order to understand the transition of the Earth System from the Holocene to the Anthropocene, $\psi$ was defined as $\psi(T):= (T -T_H)/T_H$. where $T_H$ is the average Holocene temperature. For a general planet classification, it is convenient to keep a similar definition, considering $T_{\rm t}$ as the average temperature of a \emph{temperate} Holocene-like equilibrium state of a planet. In this classification scheme, each planet would have its own temperature state, with an equilibrium temperature for the symetric phase, $T_{\rm t}$, determined by its energy balance relative to the mother star.

We can introduce more terms of powers of $\psi$ in the free energy function so to describe a richer set of equilibrium states available for a planet. A rich lore of states can be found, for instance, throughout Earth's history, however, these correspond in any case to states hotter or cooler than the Holocene and are not, in essence, too different from the states we predict from the simplest version of our model. Thus, we keep our present description as simple as possible, although sufficiently diverse to capture the most conspicuous features of rocky type planets.

Notice that odd power terms can also be included in the free energy. A linear term is used to describe the impact of human activities \cite{OB-FF1,OB-FF2,OB-FF3,OB-FF4} and cubic terms are suitable, for instance, to model metastability \cite{Paramos:2003}. These are the next natural extensions to the model to consider.

We start by introducing a cubic term, as follows:
\begin{equation}
F(\eta) = F_0 + a(\eta)\psi^2 + b(\eta)\psi^4 + c(\eta)|\psi|^3 + \ldots,
\label{eq:FreeEnergy3terms}
\end{equation}
We now take $a(\eta)$, $b(\eta)$ and $c(\eta)$ as generic functions of the natural parameters $\eta$, without making any assumption on the dependence.

Keeping the assumption that $b(\eta)$ is positive, which is necessary for there to be any equilibrium state, then for $a(\eta) < 0$ the free energy function will still exhibit the ``asymmetric'' phase with two stable minima, one corresponding to a hot state and the other to a cold state.

\begin{figure*}
    \centering
    \includegraphics[width=\textwidth]{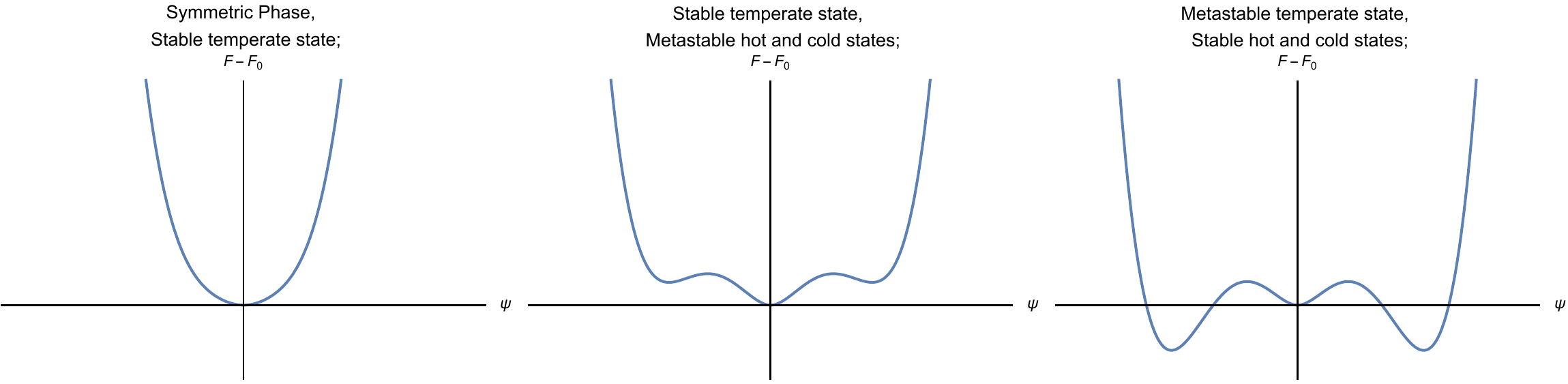}
    \caption{Free energy function for three possible phases with $a(\eta) > 0$. The left panel shows a ``symmetric'' phase with a single stable temperate state, the center panel is a phase with a stable temperate state and metastable hot and cold states, and the right panel shows a phase with stable hot and cold states and a metastable temperate state.}
    \label{fig:FreeEnergyPhases}
\end{figure*}

However, if $a(\eta) > 0$, depending on the value of the cubic term coefficient $c(\eta)$, there are three possible phases, the ``symmetric'' phase with a single stable minimum at the temperate state, a phase with a stable temperate state and metastable hot and cold states, and a phase with stable hot and cold states and metastable temperate state. These three distinct phases are depicted in Fig.~\ref{fig:FreeEnergyPhases} and summarised in Table\,\ref{tbl:FreeEnergyPhases}.

The key parameter controlling the existence of metastable states is, of course, $c(\eta)$. By solving the equation
\begin{equation}
\frac{dF}{d\psi} = 0,
\end{equation}
we find that metastable states occur for
\begin{equation}
    c(\eta) < -\frac{4}{3} \sqrt{2 a(\eta) b(\eta)}.
\end{equation}
The temperate state at $\psi = 0$ will be stable if, additionally,
\begin{equation}
c(\eta) > -2 \sqrt{2 a(\eta) b(\eta)}
\end{equation}
and metastable otherwise. In the later case there will be two stable states, one hot and one cold.

\begin{table*}
    \centering
    \caption{Summary of phases available with a free energy function given by Eq.\,(\ref{eq:FreeEnergy3terms}), as also illustrated in Fig.\,\ref{fig:FreeEnergyPhases}, with limits on $a(\eta)$, $b(\eta)$ and $c(\eta)$ for each phase.}
    {\renewcommand{\arraystretch}{1.25} %<- modify value to suit your needs
    \begin{tabular*}{\textwidth}{l | c | c | c | l}
        Phase type              & Minima  & Temperate state   & Hot and cold states   & Conditions \\
        \hline
        \hline
        Symmetric     & 1       & Stable            & None                  & $a \geq 0, ~ b > 0, ~ c > - \frac{4}{3}\sqrt{2 a b}$ \\
        Asymmetric & 3 & Stable   & Metastable & $a > 0, ~ b > 0, ~ - 2 \sqrt{ab} < c < - \frac{4}{3}\sqrt{2 a b}$ \\
        Asymmetric & 3 & Metastable   & Stable & $a > 0, ~ b > 0, ~ c <- 2 \sqrt{ab}$ \\
    \end{tabular*}}
    \label{tbl:FreeEnergyPhases}
\end{table*}

%-----------------------------------------------------------------------------%
%-----------------------------------------------------------------------------%

\section{Planetary Classification Scheme}

The physical model described in the previous section suggests a classification system for rocky planets located in or near a star's habitable zone. In fact, taking our Solar System as a reference, our scheme allows for distinguishing the three main equilibrium states: a temperate state similar to Earth during the Holocene, a hot Venus-like overheated state, and a Mars-like cold and almost globally inert state. All three planets underwent multiple changes along their respective histories. There is evidence that both Mars and Venus has been in a temperate state in the past, whereas Earth has had both cold and hot states throughout the geological record.

The hypothesis that we put forward is that these different states correspond to different phases, as modelled by the Free Energy equation, Eq. (\ref{eq:FreeEnergy3terms}). The phases and phase transitions result from the forcing due to natural causes, modelled here by the functions $a(\eta)$, $b(\eta), c(\eta), \ldots$ of $\eta$. The parameter $\eta$ itself should be the result of a wide array of physical variables.

Hence, it is essential identifying the relevant physical variables that determine the evolution of a given planet from its planetesimal origin. Planetesimals are solid proto-planets arising from the accumulation of orbiting bodies whose internal strength are dominated by self-gravity and whose orbital dynamics are not affected by the drag of gas. They evolved gravitationally in an accretion disk of the matter around a star, as hypothesised by Safranov in 1968 \cite{Safranov1968}. From this generally accepted picture and some supporting observations \cite{BlumWurm2008,HarringtonVillard2014}, we can consider as the final state of this initial development, a planet around which one can expect its own disk and a regular radiation supply from its mother star. At this phase, for a stable mother star, it is essential for the stability of a planet that the main part of its disk is already settled in stable orbits in a process dubbed ``neighbourhood clearing''. 

At least three criteria were presented to gauge the efficiency of this "cleaning" process: the Stern-Levinson parameter \cite{SL2002}, $\Lambda$; Soter's $\mu$ parameter \cite{Soter2006}; and Margot's parameter $\Pi$, \cite{Margot2015}. For further development, geophysical and biological processes are relevant, but we shall neglect in our analysis the latter. The geophysical parameters to be considered are the volcanism, $\chi$, and the existence of plate tectonics, $\phi$. Of course, a quite relevant point for considering the effect of the above mentioned parameters is to select a set of benchmark values so to track the behaviour of the $\eta$ parameter. It is natural to consider for reference either Earth values or values corresponding to stable periods of evolution of each planet. As the latter might be harder to implement in a first analysis, we shall use, whenever feasible, the Earth values for reference of the tipping points in behaviour.  

In what concerns the neighbourhood cleaning, in terms of the mass of the planet, $M_P$, the mass of its star in solar masses, $M_*$, the planet semi-major axis, $a$, the Stern-Levinson parameter is defined as:  
\begin{equation}
    \Lambda = k \frac{M_P^2}{a^{3/2}},
\end{equation}
where $k$ is a convenient constant so that $\Lambda >1$ for planets and $\Lambda <1$ for asteroids. For reference, $\Lambda_{Earth}  =1.53 \times 10^5$, $\Lambda_{Venus}  =1.66 \times 10^5$ and $\Lambda_{Mars}  =9.42 \times 10^2$.

Soter's parameter is given by the ratio of $m$ over the mass, $m_M$ of an object that shares the orbit of the candidate planet:
\begin{equation}
    \mu = \frac{M_P}{m_M},
    \label{eq:Soter}
\end{equation}
typically a candidate planet satisfies the condition $\mu >10^2$, where $\mu_{\rm Earth} =1.7 \times 10^6$, $\mu_{\rm Venus}  =1.3 \times 10^6$ and $\mu_{\rm Mars}  =5.1 \times 10^3$.

Margot's parameter is defined as
\begin{equation}
\Pi= k_{*} \frac{M_P}{M_*^{5/2} a^{9/8}},
\end{equation}
where the constant $k_{*}$ is chosen so that $\Pi >1$ for a planet that has cleared its orbital zone. Thus, the suggested discriminant for cleared orbits is $\Pi >1$. Reference values are: $\Pi_{\rm Earth}  =8.1 \times 10^2$, $\Pi_{\rm Venus}  =9.5 \times 10^2$ and $\Pi_{\rm Mars}  =5.4 \times 10^1$.

When the condition of neighbourhood cleaning is verified, another very important driver of transformations are the geological processes. For the volcanism, which triggers a series of transformations to the atmosphere, surface features and onset of oceans, we consider a conceptual benchmark value corresponding to the volcanic activity on Earth at the Holocene, say $\chi_H$.

As for plate tectonics, a particularly complex issue, opinions range from the somewhat simple view that it is controlled essentially by the mass of the planet \cite{Valencia2007}, even though it has been argued that the composition dependence should be necessarily considered, most particularly the condition for the ratio $Si/Fe>0.9$ \cite{Unterborn2016}. From the first criterion follows that super-Earth planets should have plate tectonics, while Venus should be a monoplate, which seems to be consistent with evidence from the recent observation of the Mead impact of the thickness of its lithosphere \cite{BjonnesJohnsonEvans2021} and the absence of water. However, the second criterion is not as that sharp given that $Si/Fe|_{Venus}=1.2$ \cite{Unterborn2016} and $Si/Fe|_{Mars}=0.58$ \cite{ZharkovGudkova2016}. Furthermore, given the impossibility of observing the volcanic activity of exoplanets, we shall consider instead the mass criterion. 

Of course, one key parameter that needs to be included is the variation of the energy output of the mother star, available for the planet, $L_{*}$, with respect to an average value $\bar{L_{*}}$, either due to variation of orbital parameters or due to intrinsic changes on the dynamics of the star. 

Many other parameters can be considered, but we shall keep here only the most essential ones, for the sake of exemplifying the approach. For the orbit clearing, we pick the Soter's discriminant. The volcanic activity and the energy available to the planet are clearly destabilising factors, while plate tectonics tends to help planets to reach a long term stability. Thus, we can write the 
$\eta$ parameter as
\begin{equation}
    \begin{split}
        \eta =~& \alpha \frac{(\mu-\mu_{Mars})}{\mu_{Earth}} + + \beta \frac{(M_P-M_{Earth})}{M_{Earth}} + \\
        &  + \gamma \frac{(L_{*}-\bar{L_{*}})}{\bar{L_{*}}} + \ldots,
    \end{split}
    \label{eq:q} 
\end{equation}
where $\alpha$, $\beta$ and $\gamma$ are the constants that set the relative weight of the considered factors affecting the dynamics of each planet.

Other possible parameters that could be considered are the ones associated with the presence of liquid water, of a biosphere, or the concentration of greenhouse gases.  The last set of parameters are intimately related with the issue of habitability and the long term conditions to support a stable climatic system. A recent simulation in which thousands of planets were assigned randomly generated climate feedbacks to verify whether habitable conditions remained over a period of 3 billion years showed that chance plays a role, meaning that Earth-like long lasting habitability is most likely a contingent rather than an inevitable outcome \cite{Tyrrell}.  

Clearly, an orbit clear from debris and the existence of plate tectonics are stabilising factor, while the volcanic activity above a certain threshold as well as a variation of the energy available above an average are both destabilising factors. The effect of a biosphere and more complex interactions between the various components of the planet system might require a more involved parametrisation. However, as discussed in the literature, for the existence of a biosphere, the presence of plate tectonics activity is often mentioned as a necessary condition and so our classification scheme contains the main features for a subsequent conceptual development. Another relevant feature for a biosphere is the presence of a stable and robust magnetic field; thus, a magnetic field above a certain threshold is a necessary condition for the onset and the development of a biosphere.  

Our description is generally supported by a set of observable conditions. For instance, a recent study suggests that a cloud greenhouse effect could sustain a warm ancient Mars, which is consistent with geologic data and can account for the patchy presence of surface water \cite{Kite}. This supports the idea that even a cold and arid planet as Mars at present could be the result of a transformation triggered by the change of one or a set of parameters that induced overall drastic changes from a Holocene-like conditions, as suggested by our model. The set of transformations that induced the transition of Mars towards its present state is still debatable, but there is consensus about the lack of of tectonic activity that prevented the recycling of gases locked up in sediments \cite{Fogg} and possible disappearance of a planet wide magnetosphere, as today Mars retains a magnetosphere that covers only about $40\%$ of its surface. Venus, on its hand, seems to have undergone a severe process of accumulation of greenhouse gases, which greatly increased its surface temperature to about $450^o$C and the pressure of its dense carbon dioxide atmosphere to $9~{\rm MPa}$.

Of course, our proposal of classification can be regarded as useful for exoplanets only if we can related the above discussion with known exoplanets observables. We shall discuss these issues in the next section.

%-----------------------------------------------------------------------------%
%-----------------------------------------------------------------------------%

\section{Experimental Verification and Relation with Observables} 

From the above discussion, we can conclude that a great deal of information concerning the type of a planet can be inferred from its mass, insolation and the presence of geological activity. In what follows we shall discuss how to obtain these observables and how they can be used to experimentally verify our proposed model.

We assume that a given planet that gravitates around a star of radius $R_*$, mass, $M_*$, and temperature, $T_*$, whose luminosity can inferred from the sun's radius, temperature and luminosity through the equation (see, for instance, Ref. \cite{Perryman}): 
\begin{equation}
    L_*= \left(\frac{R_*}{R_{\rm sun}}\right)^2\left(\frac{T_*}{T_{\rm sun}}\right)^4 L_{sun}.
    \label{eq:Luminosity} 
\end{equation}
    
Its rocky nature will be inferred from its density and the verification that it is located within the habitable zone is made against the limits in astronomical units (AU) set by \cite{Kasting}:
\begin{equation}
    R_{\rm HZ,inner}= 0.75 \sqrt{L_*/L_{\rm sun}} ~{\rm AU},
    \label{eq:RHinner} 
\end{equation}
\begin{equation}
R_{\rm HZ,outer}= 1.77 \sqrt{L_*/L_{\rm sun}} ~{\rm AU}.
\label{eq:RHouter} 
\end{equation}

The semi-major axis of its orbit, $a$, which is found from its period, $P$, from Kepler's third law:

\begin{equation}
a =  \left[\frac{M_*}{M_{sun}} \left(\frac{P}{yrs}\right)^2\right]^{1/3}~AU.
\label{eq:Kepler} 
\end{equation}

With the semi-major axis and star luminosity, we can obtain the insolation flux, $S$, in terms of Earth's insolation flux, $S_{Earth}$:

\begin{equation}
    S= \left(\frac{(L_*}{L_{sun}}\right)\left(\frac{AU}{a}\right)^2 S_{Earth},
    \label{eq:ins} 
\end{equation}
where for terms of comparison we should consider $S_{Venus}=1.92$, $S_{Earth}$ and $S_{Mars}=0.43 S_{Earth}$.

Once the direct or indirect observation of a planet's surface temperature becomes possible, data will start to accumulate relating surface temperatures with the insolation flux. This will be the key experimental test to our model. If this model is correct, we expect to observe the equilibrium temperatures of exoplanets in the habitable zone to accumulate around a few values corresponding to the possible phases. The number of equilibrium points, if more than three, will determine the necessity of introducing further terms to the free energy, as in Eq.\,(\ref{eq:FreeEnergy3terms}).

WIth the semi-major axis of the orbit of an exoplanet, its mass and the radius can be determined. The mass can be inferred from the radial velocity and the astrometric amplitude methods. Indeed, the radial velocity semi-amplitude, $K$, in $ms^{-1}$, can be evaluated from the planet's period, mass, $M_P$, stellar mass, $M_*$, the orbital inclination, $i$, and orbital eccentricity, $e$ through the relationship \cite{Perryman}:
\begin{equation}
    \begin{split}
        &\frac{K}{(ms^{-1})} =  203 \left(\frac{P}{\rm days}\right)^{1/3} \times \\
        & \frac{(M_P/M_J) \sin i}{((M_*/M_{\rm sun}) + 9.548 \times 10^{-4} (M_P/M_J))^{2/3}} \frac{1}{\sqrt{1-e^2}},
    \end{split}
    \label{eq:K} 
\end{equation}
where zero eccentricity and an inclination of $90^{o}$ are assumed for the ``Earth-like'' and ``Jupiter-like'' input values.

The mass be also inferred from the astrometric amplitude of the wobble of a host star induced by its companion.  It can be derived from balance of the star/planet system about its center of mass. The distance to the system then determines the angular size of the projected motion on the sky. Assuming a circular orbit, the host star, at a distance, $d$, in parsecs, describes an ellipse on the sky whose angular semi-major axis, $\Delta\theta$, is given by \cite{Perryman}:
\begin{equation}
    \frac{\Delta\theta}{\mu as} =  954.3 \frac{M_P/M_J}{M_*/M_{\rm sun}} \frac{a/AU}{d/pc},
    \label{eq:wobble} 
\end{equation}
where the numerical coefficients reflect the choice of units.

Finally, the transit depth method, allows for obtaining the radius of the planet.  Indeed, the predicted transit depth, $\delta$, is given by the ratio of the projected area of the planet to that of the star. In percent, one has \cite{Perryman} for $0<R_P<R_*$:
\begin{equation}
    \delta(\%) = 1.049 \left(\frac{R_P/R_J}{R_*/R_{sun}}\right)^2.
    \label{eq:transitdepth} 
\end{equation}

Regarding the existence of geological activity, such as volcanism and plate tectonics, the capability to directly observe its signature already exists, based on the presence of atmospheric aerosols \cite{Misra}. Furthermore, internal heating rate estimates based on tidal abd radiogenic sources can be used to infer the rate of volcanic activity of an exoplanet and the possibility of tectonic activity \cite{Lynnae}.

%-----------------------------------------------------------------------------%
%-----------------------------------------------------------------------------%

\section{Discussion and Outlook}

In this work we have proposed a classification scheme for rocky planets. Our methodology is based on the Landau-Ginsburg theory of phase transitions given its generality and that the underlying processes of transformation of planets since their origin seem to be fully captured by tracking the change of their free energy in terms of the relevant physical parameters. We have previously used this formalism to model the Holocene-Anthropocene transition of the Earth System and its evolution towards a hotter equilibrium state. In here we have identified a minimal set of parameters to describe the evolution of the planets: the absence of debris on their orbit; volcanism, plate tectonics and regular energy supply from the mother star. Of course, these parameters are a host of many other conditions. For instance, plate tectonics can be seem as a proxy for the presence of a hydrosphere and as a condition for the emergence of life. Our model can account for these complexities, but we discussed here only its simplest implementation. 

Given the simplifying assumptions discussed above and assuming that the free energy is written as up to a fourth power of the order parameter, $\psi$, we can identity three major classes of planets: Earth-like in the its Holocene-like temperate state at a stable temperature $T_{\rm t}$; a cooler ($T<T_{\rm t}$) Mars-like planet in a globally inert state; and a hotter ($T>T_{\rm t}$) Venus-like planet. The onset of the processes leading to a planet to be driven from one state into another depends of the dynamics of astronomical, geological and physical relevant properties which are triggered after crossing some  transitional benchmark value. The exact value of the temperature of the Holecene-like state is not relevant for the purpose of classification and Earth's temperature, $T_H \simeq 15^oC$ can be used as a reference value. We stress that our proposal aims to establish to which class a given planet belongs and how it might evolve once the quantity $eta$, Eq. (\ref{eq:q}), is evaluated. This can be achieved without the knowledge of the temperature $T_{\rm t}$. This is particularly relevant as the temperature is not yet an exoplanet observable. As discussed above, a positive $\eta$ value suggests that the planet is Earth-like. Negative values of $\eta$ that planet is Venus-like of Mars-like. Decision between these two possibilities can be made, for instance, through comparison with the insolation of Venus and Mars values indicated above or the density when this is available. 

The choice of Earth's Holocene values to characterise a stable equilibrium state is fairly natural, even though somewhat arbitrary. We believe that our conclusions would not be too different if another set of criteria were chosen. Indeed, this suggests that a set of surface parameters should be considered for assessing the stability of a planet state. At its present state, the Earth System is characterised by 9 parameters, the so-called Planetary Boundaries (PB)) \cite{Rockstrom}, Thus, depending on the stage, relative dimension, and importance of the components of its system, atmosphere, cryosphere, hydrosphere and upper lithosphere, it is logical to assign a set of parameters analogous to Earth's PB. Even though the relevant set of parameters, let us call them Exo-Planetary Boundaries (EPB), to consider may require a specific study and/or an i{\it in situ} analysis, it is reasonable to assume it will necessarily involve the relative importance of the mother star gravity and energy transfer and its surface features. In any case, without any further specific and detailed analysis, one can broadly state that about a third all rocky planets will be in one of our three classes: Mars-like; Earth-like; Venus-like.  

Another issue that is worth mentioning concerns the similarity between Earth and Venus and how subtle is the difference between their physical parameters. This resemblance stresses the likelihood that the Anthropocene is a transition between the Holocene to a much hotter Venus-like state. The dynamical system analysis of the Anthropocene equation emerging from our model \cite{OB-FF1} does confirm that this hotter Venus-like state is indeed an attractor of trajectories \cite{OB-FF2}. These considerations remind a speculation of 1961 of Carl Sagan \cite{Sagan1} concerning a putative terraforming or planetary engineering of Venus by seeding its atmosphere with algae to absorb its excess of $CO_2$. Subsequent considerations about the terraforming of Mars were put forward by Sagan \cite{Sagan2}, MacKay \cite{MacKay}, Lovelock and Allaby \cite{Lovelock} and others about the planetary engineering of Mercury, Moon and other planets. However, in the context of our work, terraforming is quite similar to the inverse transition of the Holocene towards a Hot House Earth scenario through the Anthropocene. This transition is modelled through a linear term in the free energy, Eq. (\ref{eq:FreeEnergy}), of the form $h(\eta)H\psi$, where $H$, models the impact of the human activities on the Earth System. It is argued that this variable can be suitably modelled at present through the PB  with respect its Holocene values plus interaction terms \cite{OB-FF2, OB-FF3}:
\begin{equation}
    H = \sum_{i=1}^{9} h_i + \sum_{i,j=1}^{9} g_{ij} h_i h_j + \sum_{i,j,k=1}^{9} \alpha_{ijk} h_i h_j h_k + \ldots,
\end{equation}
where the second and third set of terms indicate the interaction between the various effects of the human action on the planetary boundary parameters. Hence, it follows that through a suitable adaptation to the relevant EPB, one could device, likewise for preventing Earth to reach a new hotter equilibrium state, strategies to terraform the surface conditions of a given planet. This opens quite interesting possibilities for analysis. 

\vspace{0.5cm}

%{\bf Acknowledgments~~}

\noindent
%This work is partially supported by Funda\c{c}\~ao para a Ci\^encia e a
%Tecnologia (Portugal) under the project POCI/FIS/56093/2004.

%\vspace{0.3cm}

%\vfill
%\newpage

\bibliographystyle{unsrtnat}

\end{document}